# A Giant Protocluster of Galaxies at Redshift 5.7

Linhua Jiang[1], Jin Wu[1,2], Fuyan Bian[3], Yi-Kuan Chiang[4], Luis C. Ho[1,2], Yue Shen[5,6], Zhen-Ya Zheng[7,8,9], John I. Bailey, III[10,11], Guillermo A. Blanc[12,13], Jeffrey D. Crane[12], Xiaohui Fan[14], Mario Mateo[10], Edward W. Olszewski[14], Grecco A. Oyarzún[13], Ran Wang[1], Xue-Bing Wu[1,2]

[1]*Kavli Institute for Astronomy and Astrophysics, Peking University, Beijing 100871, China*

[2]*Department of Astronomy, School of Physics, Peking University, Beijing 100871, China*

[3]*Research School of Astronomy and Astrophysics, Australian National University, Weston Creek, ACT 2611, Australia*

[4]*Department of Physics & Astronomy, The Johns Hopkins University, 3400 N. Charles St., Baltimore, MD 21218, USA*

[5]*Department of Astronomy, University of Illinois at Urbana-Champaign, Urbana, IL 61801, USA*

[6]*National Centre for Supercomputing Applications, University of Illinois at Urbana-Champaign, Urbana, IL 61801, USA*

[7]*CAS Key Laboratory for Research in Galaxies and Cosmology, Shanghai Astronomical Observatory, Shanghai 200030, China*

[8]*Institute of Astrophysics and Center for Astroengineering, Pontificia Universidad Catolica de Chile, Santiago 7820436, Chile*

[9]*Chinese Academy of Sciences South America Center for Astronomy, Santiago 7591245, Chile*

[10]*Department of Astronomy, University of Michigan, Ann Arbor, MI 48109, USA*

[11]*Leiden Observatory, Leiden University, P.O. Box 9513, 2300RA Leiden, The Netherlands*

[12]*Observatories of the Carnegie Institution for Science, 813 Santa Barbara Street, Pasadena, CA 91101, USA*

[13]*Departamento de Astronomía, Universidad de Chile, Camino del Observatorio 1515, Las Condes, Santiago 7591245, Chile*

[14]*Steward Observatory, University of Arizona, 933 North Cherry Avenue, Tucson, AZ 85721, USA*

**Galaxy clusters trace the largest structures of the Universe and provide ideal laboratories for studying galaxy evolution and cosmology[1,2]. Clusters with extended X-ray emission have been discovered at redshifts up to $z \approx 2.5$[3-7]. Meanwhile, there has been growing interest in hunting for protoclusters, the progenitors of clusters, at higher redshifts[8-14]. It is, however, very challenging to find the largest protoclusters at early times when they start to assemble. Here we report a giant protocluster of galaxies at**

**redshift $z \approx 5.7$, when the Universe was only one billion years old. This protocluster occupies a volume of about $35^3$ cubic co-moving megaparsecs (cMpc$^3$). It is embedded in an even larger overdense region with at least 41 spectroscopically confirmed, luminous Lyα-emitting galaxies (Lyα Emitters, or LAEs), including several previously reported LAEs[9]. Its LAE density is 6.6 times the average density at $z \approx 5.7$. It is the only one of its kind in a LAE survey in four square degrees on the sky. Such a large structure is also rarely seen in current cosmological simulations. This protocluster will collapse into a galaxy cluster with a mass of $(3.6 \pm 0.9) \times 10^{15}$ solar masses ($M_\odot$), comparable to those of the most massive clusters or protoclusters known to date.**

According to cosmological simulations, the largest protoclusters of galaxies extend over tens of co-moving megaparsecs at $z > 5$ (refs. 15,16). Deep, wide-area surveys are needed to find these giant structures at high redshift. We are carrying out a spectroscopic survey of galaxies in four square degrees on the sky, aiming to build a large and homogeneous sample of LAEs at $z \approx 5.7$ and 6.5. We are observing five well-studied fields, including the Subaru XMM-Newton Deep Survey (SXDS) field[17]. These fields have deep optical imaging data in a series of broad and narrow bands, taken by the prime-focus imager Suprime-Cam on the 8.2-m Subaru telescope. The combined SXDS images in five broad-band filters (*BVRi'z'*) and two narrow-band filters (NB816 and NB921) have enabled us to efficiently select LAE candidates at $z \approx 5.7$ and 6.5 via the Lyα technique[18-21]. From these LAE candidates, we identified a large overdense region at $z \approx 5.7$ in the SXDS. Here we show that this overdense region contains a giant protocluster (SXDS_gPC for short) that will grow into a massive galaxy cluster.

We carried out deep spectroscopic observations of SXDS_gPC using the fiber-fed, multi-object spectrograph M2FS[22] on the 6.5-m Magellan Clay telescope. M2FS has 256 optical fibers deployed over a circular field-of-view 29.2 arcminutes in diameter. SXDS_gPC and its surrounding environment were covered by one M2FS pointing that included $z \approx 5.7$ LAE candidates brighter than NB816 = 25.8 mag (5σ detection for point sources; magnitudes are on the AB system), as well as a variety of galaxy candidates at other redshifts. We used a pair of red-sensitive gratings with a resolving power of about 2000. We obtained 7 hours of on-source integration (7 one-hour individual exposures) during dark time in November 2015. The combined spectrum reaches a Lyα flux depth of $\sim 1 \times 10^{-17}$ erg s$^{-1}$ cm$^{-2}$ (or a luminosity depth of $\sim 4 \times 10^{42}$ erg s$^{-1}$, assuming $H_0 = 68$ km s$^{-1}$ Mpc$^{-1}$, $\Omega_m = 0.3$, and $\Omega_\Lambda = 0.7$), which ensures reliable identification of LAEs down to at least NB816 = 25.5 mag. See our program overview paper[23] and Methods for details on the imaging data, target selection, and M2FS observations.

The spectroscopic observations have allowed us to remove contaminants from the LAE candidates and measure accurate redshifts for confirmed LAEs. Both are critical to characterize protoclusters. We confirmed 46 luminous LAEs at $z \approx 5.7$ from our spectroscopic data. Example spectra of four LAEs are given in Fig. 1. More details of the 46 LAEs are provided in Supplementary Fig. 1 and Table 1. Fig. 2 illustrates the SXDS field, the LAE locations, and the M2FS pointing with an areal coverage of 660 square arcminutes (arcmin$^2$). The spatial distribution of the 46 LAEs is highly uneven: 41 of them are located in the south-west half of the circular field, including some previously reported LAEs[9]. These 41 LAEs belong to the large overdense region that we identified from the photometric data. In Fig. 2, the projected area of this overdense region is enclosed by the half-circle-like shape outlined in magenta. Its size is roughly 370 arcmin$^2$, or approximates to 53 × 41 cMpc$^2$. Its real size may be larger, because the current size estimate is limited by the coverage of the Subaru imaging data and the M2FS spectroscopic data.

Figure 3 shows the redshift distribution of the LAEs. They span an effective redshift interval of $\Delta z \approx 0.10$, which corresponds to a line-of-sight depth ~ 46 cMpc. The redshift distribution implies the existence of two groups with a slight redshift offset. The majority of the LAEs are in a group at $z \approx 5.68$, with an effective redshift interval of $\Delta z \approx 0.075$, or a line-of-sight depth ~ 34 cMpc. The remaining LAEs are in a narrower redshift slice at $z \approx 5.75 \pm 0.01$. The two groups are respectively shown in blue and red in Fig. 2. The protocluster SXDS_gPC is identified from the $z \approx 5.68$ group (see Methods for details). It consists of 23 LAEs in a near-square region of 15.5 × 14.5 arcmin$^2$ (~ 37 × 35 cMpc$^2$), denoted as the cyan rectangle in Fig. 2. SXDS_gPC has a high LAE density within a giant volume ~ 35$^3$ cMpc$^3$, embedded in the even larger overdense region mentioned above. Its line-of-sight velocity dispersion is ~ 520 km s$^{-1}$, consistent with previously known large protoclusters at high redshift[12,13].

We measure galaxy overdensity $\delta_g \equiv n/\bar{n} - 1$, where n is the LAE number density in situ and $\bar{n}$ is the average density at $z \approx 5.7$. We calculate $\bar{n}$ based on the LAEs found in our M2FS survey program and the LAEs from two previous studies[20,21]. The two previous studies represent the two largest spectroscopic surveys of $z \approx 5.70$ LAEs made previously in fields other than the SXDS. They also used Suprime-Cam imaging data and the same NB816 filter for selection of LAE candidates. This substantially simplifies the calculation of $\delta_g$, because in this case a volume overdensity is equal to its corresponding surface overdensity. We count LAEs brighter than NB816 = 25.5 mag (the common magnitude limit in different samples) and correct for sample incompleteness. The resultant overdensities in the region surrounding and including SXDS_gPC are $\delta_g = 3.8 \pm 0.7$ and $\delta_g = 5.6 \pm 1.2$, respectively. This means that the LAE density in SXDS_gPC is 6.6 times the average density at $z \approx 5.7$. With the above LAEs used for the overdensity calculation, we estimate the significance of the overdensity, i.e., statistically how significant the structure is in random fields. We find that the overdensity $\delta_g = 5.6$ has a significance of ~ 5σ, indicating that the LAE overdensity in SXDS_gPC is highly significant.

We perform a test to demonstrate that SXDS_gPC is not a region with enhanced Lyα emission. We make use of the LAEs for the overdensity calculation above and estimate the fraction of LAEs brighter than 26.6 mag in the $z'$ band for each sample. The $z'$-band photometry represents UV continuum flux, because it does not cover the Lyα emission for the LAEs in this study. The adopted magnitude limit 26.6 mag is roughly the 3σ detection limit for most $z'$-band images here. We find that this fraction in SXDS_gPC is very similar to those in other fields (see Methods for details). This clearly indicates that SXDS_gPC is not a region of increased Lyα emission relative to UV continuum emission. Instead, it is a highly overdense region of galaxies. The intrinsically high overdensity of SXDS_gPC exceeds the collapse threshold considerably in the classical theory of spherical collapse. Cosmological simulations also suggest that an overdense region like SXDS_gPC will fall into a giant galaxy cluster[16].

Spectroscopically confirmed giant protoclusters like SXDS_gPC at $z > 5$ have not been reported before. We estimate how rare they are using cosmological simulations (see Methods for details). We update a previous work[16] that was based on the Millennium Run dark matter simulations[1]. We also incorporate a data-driven, semi-analytic model[24] that represents one of the latest galaxy formation models. The simulation box size is 713 cMpc on a side. We use LAEs to trace dark matter, and search for protoclusters in a cubic window (35 cMpc)$^3$ at redshift near 5.7. The window size approximates the volume of SXDS_gPC. We find no giant protoclusters like SXDS_gPC with an overdensity $\delta_g \approx 5.6$ in the entire simulation box. The highest overdensity found in the (35 cMpc)$^3$ simulation windows is about 4.4. The probability to find one system like SXDS_gPC in our survey area of four square degrees is ~ 5%, implying that such systems are rare in the distant Universe.

We use two methods to estimate the present-day mass $M_{z=0}$, which is the total mass of baryonic matter and dark matter in SXDS_gPC (see Methods for details). We first use a classic formula[8], $M_{z=0} \approx (1 + \delta_m)\bar{\rho}V$, where $\bar{\rho}$ is the current mean density of the Universe ($3.88 \times 10^{10}$ $M_\odot$ cMpc$^{-3}$), V is the volume, and $\delta_m$ is the mass overdensity. The value of $\delta_m$ is determined by $1 + b\,\delta_m = C\,(1 + \delta_g)$, where b is the bias parameter (b = 4.17 from our simulation results above) and C is a small correction for redshift-space distortion. The resultant mass is $M_{z=0} = (3.4 \pm 0.6) \times 10^{15}$ $M_\odot$. This classic approach assumes that everything within the volume will collapse into a cluster, and the mass strongly depends on the volume V. On the other hand, this mass is not very sensitive to the measured overdensity $\delta_g$ or bias b, for a given volume V.

Our second method of mass measurement is to use the correlation between galaxy overdensity and present-day mass drawn from the simulation results[16]. This method does not require that everything in a protocluster volume has to fall into a cluster, it is thus not sensitive to the collapse volume assumed. As we mentioned earlier, there are no systems like SXDS_gPC in our simulation results above, so we make use of the protocluster with the largest overdensity $\delta_g \approx 4.4$ in the simulation. This protocluster has a present-day mass of $3.3 \times 10^{15}$ $M_\odot$. Based on the classic formula above, for protoclusters with the same size at the same redshift, $M_{z=0}$ is proportional to $(1+\delta_m)$. Therefore, the mass ratio of SXDS_gPC to the $\delta_g \approx 4.4$ protocluster is the ratio of their $(1+\delta_m)$ values. From this calculation, we find that the present-day mass of SXDS_gPC is $M_{z=0} \approx (3.6 \pm 0.9) \times 10^{15}$ $M_\odot$, consistent with the mass estimated from the classic formula. The two mass measurements demonstrate that SXDS_gPC is one of the most massive clusters or protoclusters currently known[25-27].

The cold dark matter model predicts that small structures merge hierarchically to form large structures, so the largest structures are expected to form in the latest cosmic times. It is thus remarkable that giant protoclusters like SXDS_gPC already exist at $z \approx 5.7$. Although SXDS_gPC is still far from virialized, its high overdensity suggests that this large overdense region must have been in place at an even earlier time. Such protoclusters may be ideal probes for understanding early structure formation. Furthermore, the discovery of SXDS_gPC has an intriguing implication for cosmic reionization, which ended at $z \sim 6$. Some semi-analytic models and large-scale simulations have shown that the structure of reionization is mostly driven by the clustering of galaxies[28]. Under this picture, large-scale, high-density regions were ionized first, where ionizing bubbles are thought to extend over tens of co-moving megaparsecs[29,30]. The progenitor of SXDS_gPC is likely such a high-density region in the reionization era. Our results fit well into this scenario, and may provide direct evidence for the existence of large-scale clustering required by the above reionization theory.

**Acknowledgements** We acknowledge support from the National Key R&D Program of China (2016YFA0400703 and 2016YFA0400702), and from the National Science Foundation of China (grant 11533001). G.B. is supported by CONICYT/FONDECYT, Programa de Iniciacion, Folio 11150220. E.O. acknowledges support from the NSF from grant AST1313006. We thank R. de Grijs and M. Ouchi for useful discussions. This paper includes data gathered with the 6.5 meter Magellan Telescopes located at Las Campanas Observatory, Chile. Australian access to the Magellan Telescopes was supported through the National Collaborative Research Infrastructure Strategy of the Australian Federal Government.

**Author Contributions** L.J. is the Principal Investigator of the project, analysed the data, and prepared the manuscript. J.W. reduced the M2FS images. F.B., Y.S., Z.Z., J.I.B., J.D.C., M.M., and E.W.O. helped with the M2FS observations. Y.C. carried out the simulations. L.C.H., X.F., R.W., and X.W. prepared the manuscript. G.A.B. and G.A.O. helped with the M2FS data reduction. All authors helped with the scientific interpretations and commented on the manuscript.

**Author Information** Reprints and permissions information is available at www.nature.com/reprints. The authors declare no competing financial interests. Readers are welcome to comment on the online version of the paper. Correspondence and requests for materials should be addressed to L.J. (jiangKIAA@pku.edu.cn).

**Competing Interests** The authors declare that they have no competing financial interests.

**Data Availability** The data that support the plots within this paper and other finding of this study are available from the corresponding author upon reasonable request.

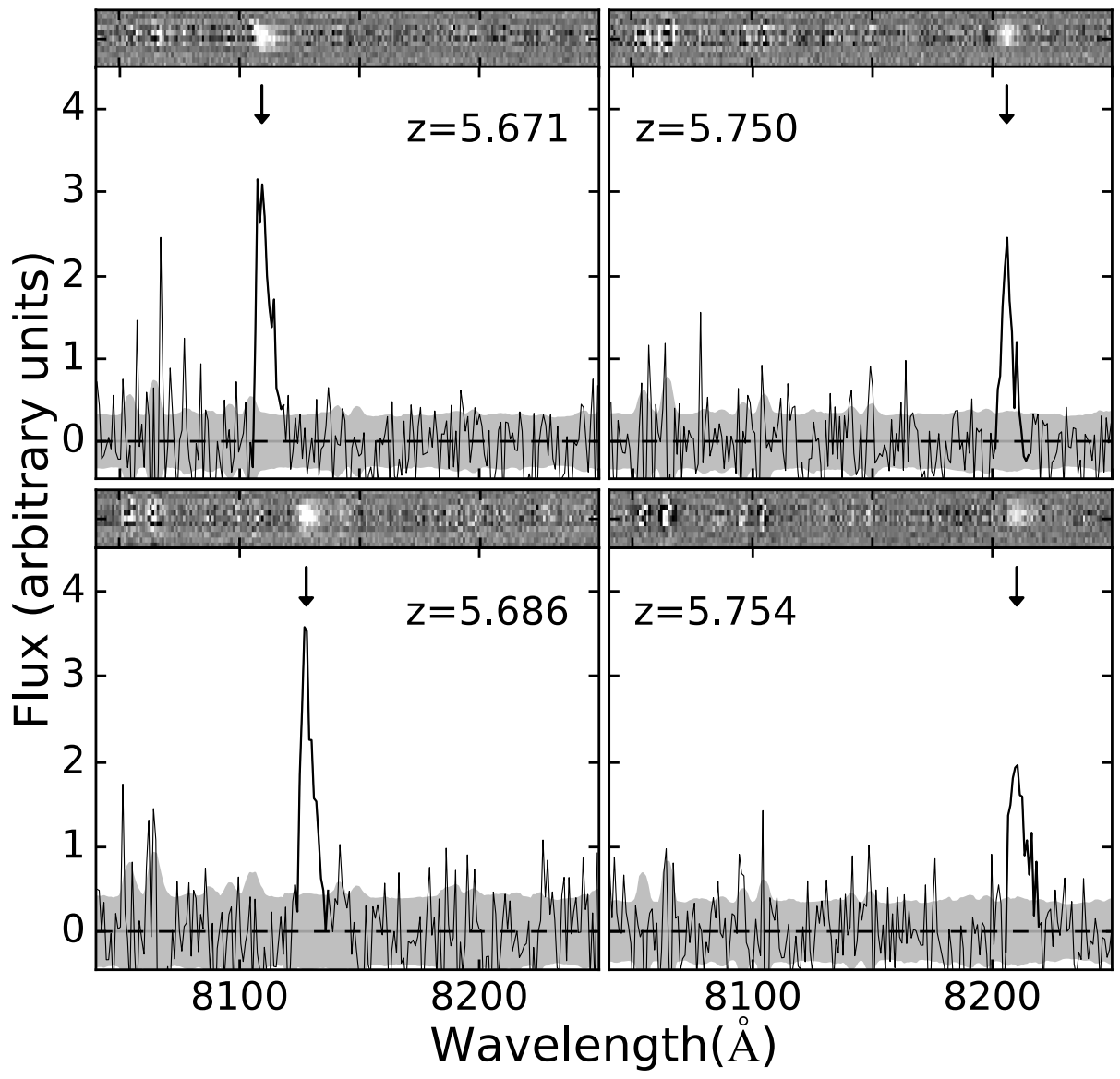

**Figure 1: Example spectra of four LAEs taken by M2FS.** In each case, we show the two-dimensional (upper) and one-dimensional (lower) spectra, with the zero-flux level (dashed line) and 1σ uncertainty region (grey region) indicated on the one-dimensional spectrum. The downward arrow points to the position of the Lyα emission line. The spectra of all 46 LAEs are shown in Supplementary Fig. 1.

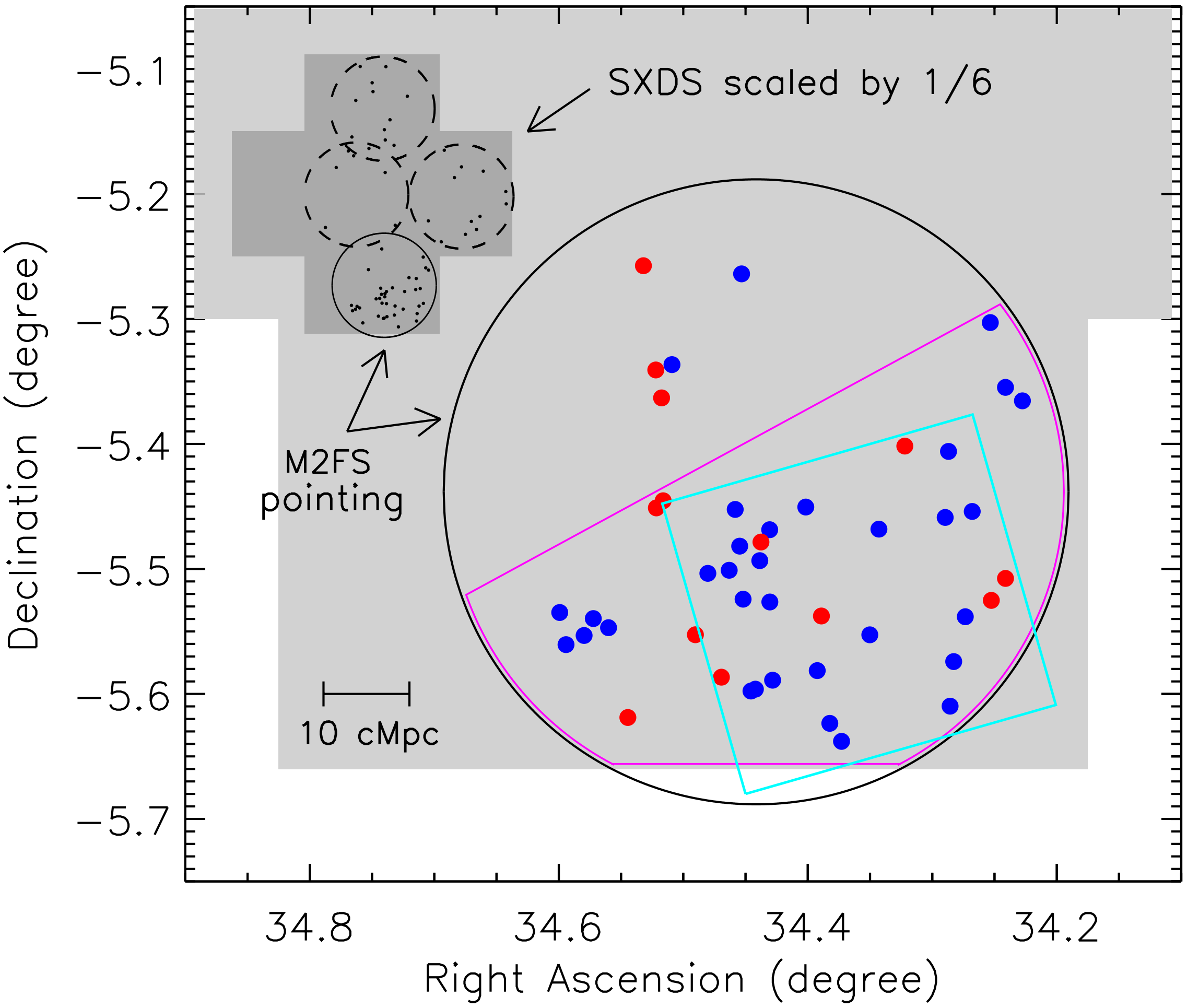

**Figure 2. Schematic representation of the SXDS_gPC region.** The light grey background shows the coverage of the Subaru imaging data, and the black circle indicates the coverage of the M2FS pointing for SXDS_gPC. The whole SXDS field, scaled by 1/6, is overplotted in dark grey in the upper-left corner, where the dashed circles plot another three M2FS pointings that have been observed by our M2FS program. The black points represent all spectroscopically confirmed LAEs brighter than NB816 = 25.5 mag, and they clearly indicate a large overdense region in the southern part of the SXDS. In the zoomed-in circle, this overdense region, with a LAE overdensity of $\delta_g \approx 3.8 \pm 0.7$, is outlined by the magenta, half-circle-like shape. The blue and red points are the 46 spectroscopically confirmed LAEs in the two groups at $z \approx 5.68$ and 5.75, respectively (see also Fig. 3). The cyan rectangle represents the giant protocluster SXDS_gPC that is embedded in the large overdense region. It consists of 23 LAEs at $z \approx 5.68$ (blue points) in a near-square region of 15.5 × 14.5 arcmin$^2$, or ~ 37 × 35 cMpc$^2$. It has a high overdensity of $\delta_g \approx 5.6 \pm 1.2$.

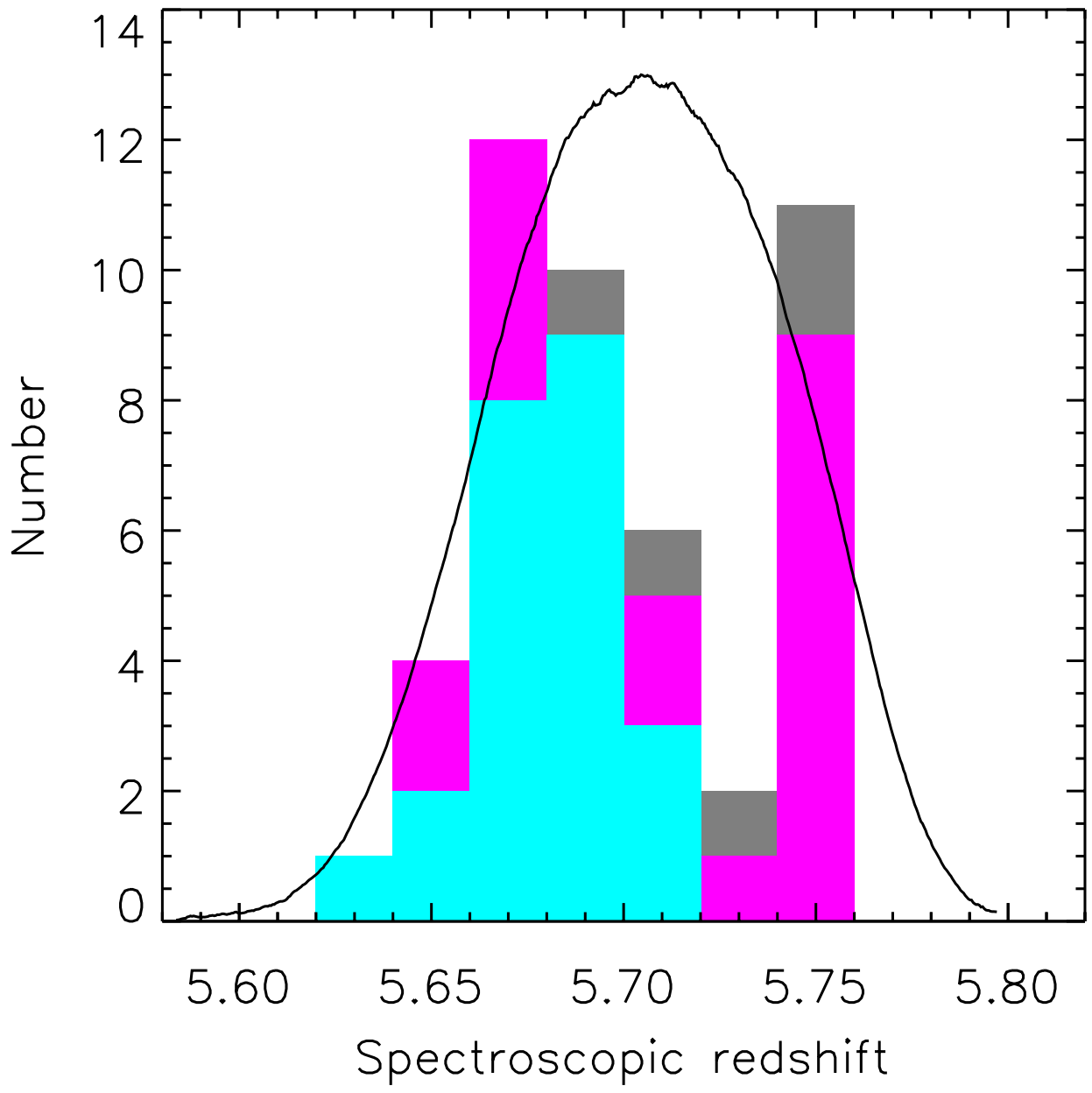

**Figure 3. Redshift distribution of the LAEs.** The dark grey, magenta, and cyan histograms include all 46 LAEs. The magenta and cyan histograms represent the 41 LAEs in the large overdense region, and the cyan histogram represents the 23 LAEs in SXDS_gPC only (see also Fig. 2). The black profile is the transmission curve of the NB816 filter used to select LAE candidates, scaled to a peak value of 13. The full width at half maximum of the filter is 120 Å, corresponding to an effective redshift interval of $\Delta z \approx 0.10$ or a line-of-sight depth of ~ 46 cMpc. The histogram shows two peaks that represent two groups at $z \approx 5.68$ and 5.75, respectively.

# METHODS

## The M2FS survey program

We are carrying out a spectroscopic survey of high-redshift galaxies in four square degrees ($deg^2$) on the sky, using the fiber-fed, multi-object spectrograph M2FS on the Magellan Clay telescope. We aim to build a large and homogeneous sample of LAEs at $z \approx$ 5.7 and 6.5, and Lyman-break galaxies at $5.5 < z < 6.8$. Our program overview paper[23] presents the details of the program, including the motivation, program design, target selection, M2FS observations, LAE identification, and scientific goals. The fields that we chose to observe are five well-studied fields, including the SXDS. These fields have deep optical imaging data in a series of broad and narrow bands, taken by the prime-focus imager Suprime-Cam on the 8.2-m Subaru telescope. The combined SXDS images in five broad bands $BVRi'z'$ have depths of 27.9, 27.6, 27.4, 27.4, and 26.2 mag, respectively. The image in the narrow band NB816 reaches a depth of 26.1 mag.

Our main targets of the M2FS program are LAEs at $z \approx 5.7$ and 6.5. We selected LAE candidates using the narrow-band (or Lyα) technique. This technique has been extensively addressed in the literature, and selection criteria in different studies are quite similar. Here we focus on the selection of $z \approx 5.7$ LAEs in the SXDS. The selection was mainly based on the $i'$ − NB816 > 1.0 color cut. We further required that candidates should not be detected ($< 2\sigma$) in any bands bluer than $R$. We visually inspected each candidate and removed spurious detections caused by diffraction spikes from bright stars, cosmic rays, and satellite trails, etc. We mainly considered relatively bright LAEs with NB816 ≤ 25.5 mag, but also included some slightly fainter candidates when possible.

Owing to a large field-of-view (FoV) and high throughput, M2FS is very efficient to identify relatively bright, high-redshift galaxies. We use a pair of red-sensitive gratings with a resolving power about 2000. The wavelength coverage is roughly from 7600 to 9600 Å. The design of the M2FS pointings or plug plates is described in the overview paper[23]. For the SXDS, we use five M2FS pointings to cover most of its area (~1.2 $deg^2$), including one pointing covering SXDS_gPC and its surrounding region. Spectroscopic observations for our M2FS program are still going on. We have observed a total of ~2.5 $deg^2$ so far. The depth is not uniform, but many M2FS pointings have reached the required depth, including the pointing with SXDS_gPC.

## LAEs in the SXDS_gPC region

The M2FS observations of the SXDS_gPC region were made during dark time in November 2015. From these observations, we confirmed 46 luminous LAEs at $z \approx 5.70 \pm 0.05$ from their Lyα emission lines. The details of the 46 LAEs are provided in Supplementary Fig. 1 and Table 1. The Lyα emission lines were identified and confirmed based on both two-dimensional (2D) images and one-dimensional (1D) spectra. Our LAE selection criteria generally ensure that a detected line in the expected wavelength range around 8160 Å is the Lyα emission line of a LAE at $z \approx 5.7$. Four strong emission lines in star-forming galaxies that might contaminate our lines are [OII] λ3727, Hβ, [OIII] λ5007, and Hα. The non-detection in very deep *BVR* images and the wavelength coverage of our spectral data rule out the possibility that the detected line is Hβ, [OIII] λ5007, or Hα. The most likely contaminants are [OII] emitters, as already pointed out in the literature. The [OII] line is a doublet, and our spectral resolution is high enough to resolve it. A tiny fraction of the candidates are indeed confirmed to be [OII] emitters at lower redshift. Finally, we can clearly see asymmetry in the emission lines of the confirmed LAEs. This is an indicator of the Lyα emission line at high redshift due to strong IGM absorption blueward of the line.

Figure 2 illustrates the layout of the SXDS_gPC region. The whole SXDS field is scaled by 1/6 and is shown in dark grey in the figure. The black points represent spectroscopically confirmed LAEs brighter than NB816 = 25.5 mag. Their spatial distribution clearly indicates the existence of a large overdense region in the southern part of the SXDS. This is the overdense region that we identified from the photometric data. We use the magenta, half-circle-like shape to denote this large overdense region. Most parts of its boundary are confined by the coverage of the imaging and spectroscopy available. Its upper (north-east) boundary is described as a straight line that is close to the two nearest LAEs within the overdense region. The projected size approximates to 22 × 17 $\mathrm{arcmin}^2$, or 53 × 41 $\mathrm{cMpc}^2$. The real size may be larger and beyond the coverage of the available data. This field was previously claimed[9] to have two compact overdense regions with eight secure LAEs at $z \sim 5.7$. Their sizes were about several co-moving megaparsecs, and their overdensities were estimated to be around 80. There has been no previous confirmation of giant protoclusters at $z \sim 5.7$ like SXDS_gPC in this field.

Figure 3 shows the redshift distribution of the LAEs and the transmission curve of the NB816 filter. The starting point of the histogram is z=5.62 and the bin size is 0.02. The full width at half maximum of the filter is 120 Å, corresponding to an effective redshift interval of $\Delta z \approx 0.10$, or a line-of-sight depth of ~ 46 cMpc. Here the line-of-sight depth derived from the redshift interval is a good approximation of the real line-of-sight distance. In the very early stage of the collapse, high-redshift protoclusters were just breaking away from the Hubble flow, so the redshift-space distortion is small.

Figure 3 clearly shows two distribution peaks at $z \approx 5.68$ and 5.75. The blue and red points in Fig. 2 represent the LAEs in the two groups, respectively. The majority of the LAEs are in the $z \approx 5.68$ group. The protocluster SXDS_gPC is identified from this group. In the literature, there was no clear definition for the boundary of a protocluster: previous studies tend to draw a box (square or rectangle) or a circle to include galaxies in a relatively isolated overdense region. Follow this convention, we use a cyan rectangle in Fig. 2 to denote the SXDS_gPC boundary. Note that we have excluded five LAEs at RA ~ 34.6 that are far away from other LAEs. We have also excluded three LAEs at Dec ~ −5.35 for the same reason. The remaining LAEs appear to be relatively isolated and have a near-square shape, based on their spatial distribution. Therefore, we use the cyan rectangle to include these LAEs as the boundary of SXDS_gPC. The distance between each side of the rectangle and its nearest LAE is the median separation between the LAEs and their nearest neighbors. The projected size of SXDS_gPC is about 15.5 × 14.5 arcmin$^2$, or ~ 37 × 35 cMpc$^2$. This definition of the boundary is conservative. If the three LAEs at Dec ~ −5.35 were included, the projected size would increase by ~20% (but the galaxy overdensity would decrease by ~5%; see the next sections). The line-of-sight depth of SXDS_gPC is ~ 34 cMpc, so the volume is about 35$^3$ cMpc$^3$. We estimate the line-of-sight velocity dispersion from the spectroscopic redshifts using a biweighted standard deviation. The measured value is about 520 km s$^{-1}$, consistent with those of other large protoclusters known at high redshift.

**The overdensity in SXDS_gPC**

Galaxy overdensity is defined as $\delta_g \equiv n/\bar{n} - 1$. We calculate the average density $\bar{n}$ based on our M2FS survey program and two previous studies[20,21]. We count LAEs brighter than NB816 = 25.5 mag, which is the common magnitude limit in different studies. In this section, LAEs are meant to be LAEs brighter than 25.5 mag. As we mentioned earlier, we have observed a total of ~2.5 deg$^2$ for our M2FS program. Although the depth is not uniform for the M2FS pointings observed so far, most pointings have reached the required depth. We find 86 $z \approx 5.70$ LAEs over a total of 5160 arcmin$^2$ in these pointings (excluding the pointing with SXDS_gPC).

The previous studies Hu et al. (2010) and Kashikawa et al. (2011) (refs. 20 and 21) represent the two largest spectroscopic surveys of $z \approx 5.70$ LAEs that were previously conducted in fields other than the SXDS. The two studies also used Suprime-Cam imaging data and the same NB816 filter for selection of $z \sim 5.70$ LAE candidates, meaning that a volume overdensity is equal to its corresponding surface overdensity because of the same line-of-sight depth. The Hu et al. sample consists of 7 individual fields, with a total coverage of 4180 arcmin$^2$. More than 90% of their LAE candidates were spectroscopically observed down to a limit of NB816 ~ 25.5 mag. They identified 73 LAEs. The Kashikawa et al. sample contains one field, namely the Subaru Deep Field (SDF; 867 arcmin$^2$). More than 90% of their LAE candidates brighter than 25.5 mag have been spectroscopically observed (see their Fig. 7). Most of the candidates were actually observed by Shimasaku et al. (2006), who confirmed 28 LAEs brighter than 25.5 mag[19].

In order to calculate the galaxy overdensity, we first correct for sample incompleteness. Sample incompleteness originates from four major sources: 1) object detection in imaging data, 2) galaxy candidate selection, 3) spectroscopic observations, and 4) LAE identification in spectroscopic data. The first completeness is the probability to detect an object in the NB816 images. The probability is roughly constant before the object brightness reaches a faint limit. The common limit for all samples used to calculate $\bar{n}$ is 25.5 mag (e.g., Fig. 3 in Hu et al. 2010). Therefore, we choose 25.5 mag as our limit, so that we do not need to apply corrections for this incompleteness.

The second source of sample incompleteness comes from target selection, i.e., the color cuts used to select targets. Different color cuts select different fractions of real LAEs. The color cut that we used is $i' -$ NB816 > 1.0. We do not compute absolute completeness here.

Instead, we apply small corrections so that each sample has the same color cut $i' - $ NB816 > 1.0 (therefore the same completeness). Hu et al. (2010) used $I -$ NB816 > 0.8. The filter $I$ is slightly redder than $i'$, and this cut is almost the same as our cut when the small difference is considered. Shimasaku et al. (2006) also used the same criterion. Kashikawa et al. (2011) used a more stringent cut $i' -$ NB816 > 1.5. We find that the fraction of LAEs with $i' -$ NB816 < 1.5 is 16% in the combination of the Hu et al. sample and our sample, so we apply a correction factor 1/0.84.

The third source for sample incompleteness comes from spectroscopic observations, i.e., the fraction of photometrically selected targets that were observed spectroscopically. This fraction is about 97% in our program. We actually included nearly 100% LAEs in our M2FS observations, but we missed ~3% of the targets due to fiber problems (such as very low-efficiency fibers). We thus apply a correction factor 1/0.97 to our sample. The Hu et al. sample and Kashikawa et al. sample have a completeness of roughly 90%, as we mentioned earlier. We apply a correction factor of 1/0.9 to the two samples.

The fourth major source for incompleteness comes from LAE identification in spectra. Because there are numerous OH skylines in the red part of the optical range, the fraction of LAEs that can be recovered from spectra is a function of wavelength for a given spectral resolution. However, the NB816 filter is located in an OH-dark window, so the effect of OH skylines is negligible. The completeness may also depend on instruments, observing modes, weather conditions, and so on. As long as the spectroscopic data are deep enough, the completeness here is nearly100%.

In summary, with the above incompleteness corrections, there are 73/0.9 LAEs over 4180 arcmin$^2$ from the Hu et al. sample, 28/0.9 LAEs over 867 arcmin$^2$ from the Kashikawa et al. sample, and 86/0.97 LAEs over 5160 arcmin$^2$ from our sample. The average surface density is 0.0197 per arcmin$^2$. In the large overdense region, we have 35 LAEs brighter than NB816 = 25.5 mag in ~370 arcmin$^2$. The density is 0.095 per arcmin$^2$, about 4.8 times the average density. The resultant overdensity is $\delta_g = 3.8 \pm 0.7$, where the error is from Poisson statistics. Similarly, we have 22 LAEs brighter than NB816 = 25.5 mag in SXDS_gPC, and the overdensity is $\delta_g = 5.6 \pm 1.2$. Note that the line-of-sight depth of SXDS_gPC is 75% of that of the large overdense region.

Using the same LAEs used for the overdensity calculation above, we estimate the significance of the overdensity in SXDS_gPC, i.e., statistically how significant the structure is compared to random fields. The Hu et al. sample consists of 7 individual fields, and the field size of 25 × 25 arcmin$^2$ is similar to the M2FS FoV. The Kashikawa et al. sample is located in one field, and its size is slightly larger. We randomly place a 15.5 × 14.5 arcmin$^2$ cell (the size of SXDS gPC) in all fields that we used above (including our own fields but excluding the SXDS_gPC region), and count LAEs in the cell. A cell is not used if it crosses an edge of a field. This is repeated one thousand times. The resultant statistics show that the overdensity of SXDS_gPC is at ~5σ significance.

**The cosmological simulation**

We use a cosmological simulation to search for protoclusters at high redshift. This is an update of a previous work[16]. The simulation has adopted the Planck cosmology, and incorporated one of the latest galaxy formation models[24], which is a data-driven, semi-analytic model. The simulation box size is 713 cMpc on a side. Galaxies selected on the basis of stellar mass or UV luminosity are often used as tracers of dark matter. But our LAE sample was not selected by stellar mass, nor by UV luminosity. Instead, it is a Lyα-flux-limited sample (limited by the NB816-band flux). Therefore, we use LAEs as tracers of dark matter, by linking Lyα emission to star formation rate (SFR).

We have followed the previous work[16] that used galaxies selected on the basis of SFR as tracers. We link Lyα emission to SFR using an empirical relation in Jiang et al. (2013) (ref. 31) between Lyα-based SFR(Lyα) and UV-based SFR(UV). We also make the common assumption that SFR(UV) is the intrinsic SFR. For our sample, the SFR(Lyα) limit is roughly 4 $M_\odot$ $yr^{-1}$, corresponding to SFR(UV) ~ 1 $M_\odot$ $yr^{-1}$ based on the empirical relation mentioned above. We further perform a test to check the limit of SFR(UV) using the $z'$-band photometry. As seen from Supplementary Table 1, many LAEs are not significantly detected in the $z'$ band. We visually identify the 9 faintest LAEs that are barely detected in the $z'$ band. We combine (average) them and do a forced photometry on the expected position. The combined $z'$-band photometry is 28.7 mag. We assume a typical rest-frame UV slope of –2 and convert the $z'$-band photometry to SFR(UV), as we did in Jiang et al. (2013). We find SFR(UV) = 0.92 $M_\odot$ $yr^{-1}$. This is well consistent with the above limit from SFR(Lyα). Therefore, we adopt 1 $M_\odot$ $yr^{-1}$ as the SFR limit of our sample.

We then use galaxies with SFR > 1 $M_\odot$ $yr^{-1}$ as tracers to search for protoclusters in the simulation. The LAE bias parameter b at $z$ ~ 5.7 derived from our simulation results is b = 4.17. This is well consistent with the bias b = 4.11 ± 0.17 from the recent results of the extensive photometric survey of $z$ ~ 5.70 LAEs over ~14 $deg^2$ from the Subaru HSC program[32]. Note that the measured present-day mass $M_{z=0}$ is not sensitive to the SFR limit that we adopt. The empirical relation between SFR(Lyα) and SFR(UV) has a scatter smaller than 50%. Even if we double the SFR limit to 2 $M_\odot$ $yr^{-1}$, the bias will slightly increase to b = 4.64 (an increase of ~10%; assuming it is an upper limit). Based on the classic formula used in the main text and the next section, $M_{z=0}$ is proportional to (1+ $\delta_m$). An increase of b from 4.17 to 4.64 results in a ~5% decrease in the estimated $M_{z=0}$.

Finally, we perform a test to show that SXDS_gPC is not simply a region of increased Lyα emission relative to UV emission. We make use of the LAEs that have been used to measure the galaxy overdensity in SXDS_gPC, and calculate the fraction of LAEs brighter than 26.62 mag in the $z'$ band in each sample. The magnitude limit of 26.62 mag adopted here is the 3σ limit for the Kashikawa et al. sample (see also Shimasaku et al. 2006). The 3σ limit of the most $z'$-band images in our M2FS program is similar to this limit. We exclude the fields that have much shallower $z'$-band images. The Hu et al. sample does not have $z'$-band photometry available. In the Kashikawa et al. sample, there are 9 out of 28 LAEs with $z'$-band detections (i.e., brighter than 26.62 mag). After we correct for a sample incompleteness due to a slightly different color select criterion (see the previous section), the fraction is (9+28×0.16)/28≈48%. Strictly speaking, this fraction is the upper limit, assuming that all missed galaxies are brighter than the limit of 26.62 mag. In SXDS_gPC, the fraction of LAEs with $z'$<26.62 mag is 57%. In the fields other than SXDS_gPC in our M2FS program (excluding shallow regions mentioned above), the fraction is 53%. We can see that this fraction in SXDS_gPC is similar to, or slightly higher than those in other fields. This clearly indicates that SXDS_gPC is not a region with enhanced Lyα emission.

**The mass of SXDS_gPC**

We estimate the probability to find one system like SXDS_gPC in our survey area of four deg$^2$, based on the number of protoclusters found in the simulation. The expected number of protoclusters similar to SXDS_gPC in terms of volume and overdensity is equal to $\int n(\delta_g) P(\delta_g)\, d\delta_g$, where $n(\delta_g)$ is the number of protoclusters with $\delta_g$ in the simulation, and $P(\delta_g)$ is the probability distribution of $\delta_g$ for SXDS_gPC. The resultant probability to find one such system in four deg$^2$ is ~ 5%. Therefore, systems like SXDS_gPC are rare at high redshift.

We have used two methods to estimate the present-day mass ($M_{z=0}$) of SXDS_gPC. The details are described in the main text. The first one is a widely used formula $M_{z=0} \approx (1+\delta_m)\bar{\rho}V$ (ref. 8). The value of $\delta_m$ is determined by $\delta_g$ via the bias parameter b and a small correction factor C for redshift-space distortion, $1 + b\, \delta_m = C\, (1 + \delta_g)$, where $C = 1 + f - f\, (1 + \delta_m)^{1/3}$ and $f = \Omega_m\, z^{4/7}$. The result is $M_{z=0} = (3.4 \pm 0.6) \times 10^{15}\ M_{\odot}$, where the error includes the uncertainties of V, b, and $\delta_g$. This classic approach assumes that everything within the volume will collapse into a cluster, so the mass strongly depends on (proportional to) the volume V. As we can see from Fig. 2, we adopt a very conservative definition of the SXDS_gPC boundary that is nearly a lower limit. We assume a 10% relative uncertainty for V. On the other hand, the mass is not sensitive to $\delta_g$ or b, for a given V. Based on this classic formula, a 20% (~1σ uncertainty) increase of $\delta_g$ results in a ~10% increase in $M_{z=0}$, and a 10% increase (upper limit, assumed to be 1σ uncertainty) of b results in a 5% decrease in $M_{z=0}$.

In the second method, we estimate the present-day mass by comparing with the simulation results based on the relation between galaxy overdensity and present-day mass[16]. We have used the $\delta_g = 4.4$ protocluster with $M_{z=0} = 3.3 \times 10^{15}\ M_{\odot}$ in the simulation. By scaling its mass, we obtain $M_{z=0} \approx (3.6 \pm 0.9) \times 10^{15}\ M_{\odot}$ for SXDS_gPC. The mass error is dominated by the uncertainty from the simulation. This uncertainty is the combination of the uncertainties that are reflected by the scatter in the relation between mass and galaxy overdensity. The mass error also includes the uncertainties of b and $\delta_g$ when we scale the mass of the $\delta_g = 4.4$ protocluster to the mass of SXDS_gPC. In addition, we assume a 10% relative uncertainty for V, as we did above. This method is based on a simulated correlation, and does not require that the whole protocluster region needs to entirely collapse. The derived mass is not sensitive to the assumed collapse volume V. For example, if we use a (45 cMpc)$^3$ window size to search for protoclusters in the simulation, the volume size is increased by a factor of 2, but the mass is only increased by ~15%.

# SUPPLEMENTARY INFORMATION

## 1. SUPPLEMENTARY TABLE 1

Supplementary Table 1 below presents the list of the 46 luminous LAEs from our M2FS observations. Columns 2 and 3 are the coordinates of the 46 LAEs. Columns 4 through 6 show their AB magnitudes in two broad bands *i′* and *z′* and one narrow band NB816 (a 2σ detection limit is provided if not detected), respectively. Column 7 lists the redshifts measured from their Lyα emission lines. The redshift errors are roughly 0.001.

**Supplementary Table 1**

| No. | $\alpha$ (J2000.0) *h m s* | $\delta$ (J2000.0) ° ′ ″ | *i′* (mag) | *z′* (mag) | NB816 (mag) | Redshift |
|---|---|---|---|---|---|---|
| 1 | 02 16 54.62 | −05 21 55.8 | 27.70±0.45 | >27.0 | 24.68±0.08 | 5.715 |
| 2 | 02 16 57.88 | −05 30 27.4 | 27.04±0.24 | >27.0 | 25.68±0.18 | 5.759 |
| 3 | 02 16 57.89 | −05 21 17.1 | 26.69±0.16 | >27.0 | 24.46±0.06 | 5.669 |
| 4 | 02 17 00.62 | −05 31 30.5 | 27.23±0.28 | 26.38±0.37 | 25.42±0.15 | 5.754 |
| 5 | 02 17 00.81 | −05 18 10.5 | >28.0 | >27.0 | 25.47±0.15 | 5.713 |
| 6 | 02 17 04.30 | −05 27 14.4 | 26.30±0.11 | 26.25±0.31 | 23.98±0.04 | 5.687 |
| 7 | 02 17 05.64 | −05 32 17.7 | 26.21±0.10 | 25.89±0.22 | 25.08±0.10 | 5.647 |
| 8 | 02 17 07.87 | −05 34 26.8 | 26.39±0.13 | 26.04±0.27 | 23.61±0.03 | 5.680 |
| 9 | 02 17 08.56 | −05 36 35.2 | 26.55±0.17 | >27.0 | 25.35±0.14 | 5.647 |
| 10 | 02 17 08.87 | −05 24 21.5 | 27.93±0.52 | >27.0 | 24.82±0.08 | 5.675 |
| 11 | 02 17 09.51 | −05 27 31.7 | 27.93±0.51 | >27.0 | 25.32±0.13 | 5.676 |
| 12 | 02 17 17.29 | −05 24 06.2 | 27.84±0.49 | >27.0 | 25.41±0.15 | 5.743 |
| 13 | 02 17 22.28 | −05 28 05.4 | 27.33±0.30 | >27.0 | 25.01±0.10 | 5.682 |
| 14 | 02 17 24.04 | −05 33 09.7 | 25.68±0.06 | 25.05±0.10 | 23.48±0.02 | 5.708 |
| 15 | 02 17 29.49 | −05 38 16.6 | 26.21±0.15 | 26.05±0.35 | 24.35±0.07 | 5.671 |
| 16 | 02 17 31.76 | −05 37 24.8 | 27.20±0.29 | 26.69±0.32 | 25.63±0.17 | 5.676 |
| 17 | 02 17 33.38 | −05 32 15.6 | 27.48±0.34 | 26.58±0.42 | 25.89±0.21 | 5.735 |
| 18 | 02 17 34.17 | −05 34 52.9 | >28.0 | >27.0 | 25.50±0.15 | 5.710 |
| 19 | 02 17 36.39 | −05 27 01.8 | 26.89±0.19 | >27.0 | 24.48±0.06 | 5.674 |
| 20 | 02 17 42.78 | −05 35 20.4 | 27.06±0.25 | >27.0 | 25.16±0.12 | 5.688 |
| 21 | 02 17 43.31 | −05 31 35.2 | 26.25±0.11 | 25.44±0.15 | 25.10±0.11 | 5.628 |
| 22 | 02 17 43.34 | −05 28 07.1 | 25.96±0.08 | 25.89±0.23 | 23.87±0.03 | 5.686 |
| 23 | 02 17 45.02 | −05 28 42.6 | 26.48±0.13 | 25.85±0.22 | 25.19±0.12 | 5.751 |
| 24 | 02 17 45.26 | −05 29 36.1 | 26.55±0.15 | 25.97±0.25 | 24.03±0.04 | 5.688 |
| 25 | 02 17 46.10 | −05 35 46.3 | 26.97±0.23 | >27.0 | 25.07±0.11 | 5.683 |
| 26 | 02 17 46.95 | −05 35 51.7 | 27.12±0.26 | 26.46±0.40 | 25.26±0.13 | 5.677 |
| 27 | 02 17 48.47 | −05 31 27.1 | 26.30±0.11 | 25.64±0.18 | 24.26±0.05 | 5.690 |
| 28 | 02 17 48.76 | −05 15 50.1 | >28.0 | >27.0 | 25.28±0.12 | 5.681 |
| 29 | 02 17 49.13 | −05 28 54.3 | 26.08±0.09 | 25.60±0.17 | 24.04±0.04 | 5.696 |
| 30 | 02 17 49.99 | −05 27 08.6 | 26.91±0.20 | 26.42±0.37 | 24.90±0.09 | 5.694 |
| 31 | 02 17 51.15 | −05 30 03.9 | 27.96±0.51 | 25.66±0.18 | 25.49±0.15 | 5.712 |
| 32 | 02 17 52.65 | −05 35 11.8 | 25.11±0.03 | 24.57±0.06 | 24.05±0.04 | 5.759 |
| 33 | 02 17 55.27 | −05 30 12.7 | 26.94±0.19 | >27.0 | 25.07±0.10 | 5.662 |
| 34 | 02 17 57.67 | −05 33 09.5 | 27.06±0.21 | 26.57±0.39 | 25.77±0.18 | 5.750 |
| 35 | 02 18 02.19 | −05 20 11.5 | >28.0 | >27.0 | 25.66±0.16 | 5.715 |
| 36 | 02 18 03.88 | −05 26 43.6 | 27.14±0.24 | 25.98±0.23 | 25.74±0.18 | 5.749 |
| 37 | 02 18 04.18 | −05 21 47.3 | 26.71±0.16 | 26.03±0.24 | 25.20±0.11 | 5.734 |
| 38 | 02 18 05.19 | −05 27 04.2 | 27.75±0.42 | 26.80±0.49 | 25.63±0.16 | 5.748 |
| 39 | 02 18 05.30 | −05 20 26.9 | 27.38±0.29 | >27.0 | 25.65±0.17 | 5.743 |
| 40 | 02 18 07.69 | −05 15 26.7 | 27.17±0.26 | 26.49±0.40 | 25.76±0.19 | 5.750 |
| 41 | 02 18 10.69 | −05 37 07.8 | 26.83±0.20 | 26.34±0.35 | 25.21±0.12 | 5.749 |
| 42 | 02 18 14.41 | −05 32 49.3 | 26.74±0.17 | 25.81±0.21 | 24.71±0.07 | 5.673 |
| 43 | 02 18 17.34 | −05 32 23.0 | 26.27±0.11 | 26.19±0.30 | 24.58±0.07 | 5.644 |
| 44 | 02 18 19.13 | −05 33 11.7 | 27.75±0.43 | >27.0 | 25.36±0.13 | 5.677 |
| 45 | 02 18 22.60 | −05 33 38.1 | 27.15±0.24 | 26.31±0.33 | 25.03±0.09 | 5.650 |
| 46 | 02 18 23.82 | −05 32 05.5 | 27.88±0.49 | >27.0 | 25.07±0.10 | 5.679 |

# 2. SUPPLEMENTARY FIGURE 1

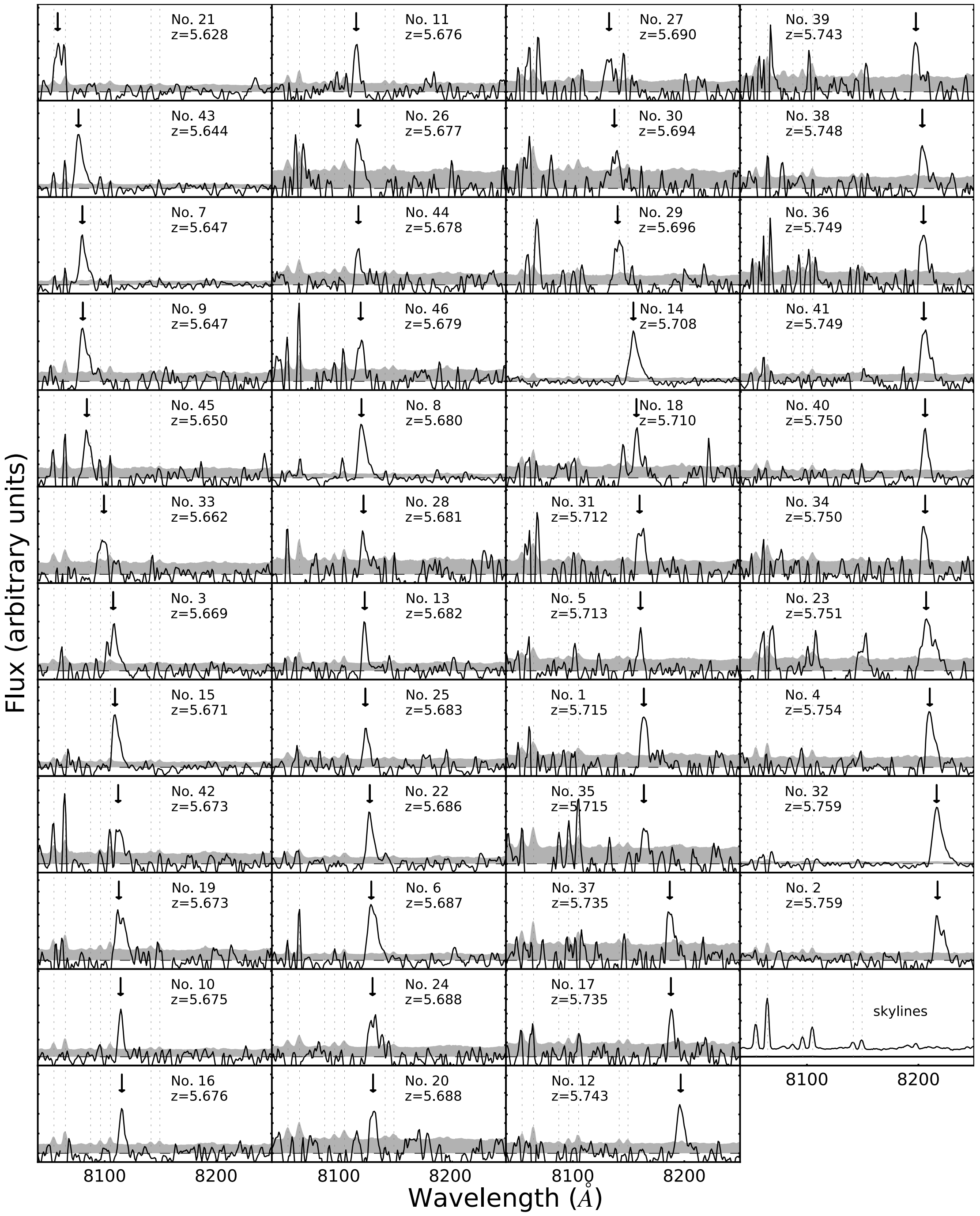

**Supplementary Figure 1. M2FS spectra of the 46 LAEs, ordered by redshift.** The spectral dispersion is ~1 Å per pixel. The spectra have been smoothed with a Gaussian kernel (a σ of one pixel is used). For each LAE, the horizontal dashed line indicates the zero-flux level, and the grey region indicates the 1σ uncertainty region. The vertical dotted lines show the positions of several OH skylines. The downward arrow points to the position of the Lyα emission line. The object number corresponds to the number in Column 1 of Supplementary Table 1.